\documentclass[pra,reprint,nofootinbib,showpacs,superscriptaddress]{revtex4-1}
\usepackage{graphicx} 
\usepackage{hyperref}
\usepackage{amsfonts}
\usepackage{amsmath,amssymb}
\usepackage{mathrsfs}
\usepackage{amsthm}
\usepackage{dsfont}
\usepackage{bm} 
\usepackage{color}
\usepackage{epstopdf} 
\usepackage{epsfig}
\usepackage{subfig} 

{\rm }

\def\be{\begin{equation}}
 \def\ee{\end{equation}}
 \def\bea{\begin{eqnarray}}
 \def\eea{\end{eqnarray}}
 \def\bes{\begin{eqnarray}}
 \def\ees{\end{eqnarray}}
 \def\bi{\begin{itemize}}
 \def\ei{\end{itemize}} 

\newcommand{\ket}[1]{|#1\rangle}

\renewcommand{\sec}[1]{\hyperref[sec:#1]{Sec.~\ref{sec:#1}}}

\newcommand{\fig}[1]{\hyperref[fig:#1]{Fig.~\ref{fig:#1}}}

\def\2{\frac{1}{2}}
\def\4{\frac{1}{4}}

\begin{document}


\title{Quantum simulation of quantum field theory using continuous variables}

\author{Kevin Marshall}
\affiliation{Department of Physics, University of Toronto, Toronto, M5S 1A7, Canada}
\author{Raphael Pooser}
\affiliation{Quantum Information Science Group, Oak Ridge National Laboratory, Oak Ridge, Tennessee 37831, U.S.A}
\affiliation{Department of Physics and Astronomy, The University of Tennessee, Knoxville, Tennessee 37996-1200, U.S.A.}
\author{George Siopsis}
\email{siopsis@tennessee.edu}
\affiliation{Department of Physics and Astronomy, The University of Tennessee, Knoxville, Tennessee 37996-1200, U.S.A.}
\author{Christian Weedbrook}
\affiliation{QKD Corp, 60 St.~George St., Toronto, M5S 1A7, Canada}

\date{\today}

\begin{abstract}

Much progress has been made in the field of quantum computing using continuous variables over the last couple of years. This includes the generation of extremely large entangled cluster states (10,000 modes, in fact) as well as a fault tolerant architecture. This has led to the point that continuous-variable quantum computing can indeed be thought of as a viable alternative for universal quantum computing. With that in mind, we present a new algorithm for continuous-variable quantum computers which gives an exponential speedup over the best known classical methods. Specifically, this relates to efficiently calculating the scattering amplitudes in scalar bosonic quantum field theory, a problem that is believed to be hard using a classical computer. Building on this, we give an experimental implementation based on cluster states that is feasible with today's technology.
\end{abstract}

\maketitle

\section{Introduction}

For more than a decade now, continuous-variable (CV) quantum information~\cite{Braunstein2005,Weedbrook2011} has been a prominent substrate in implementing quantum technologies. Primarily this can be attributed to its largely Gaussian nature which invites simple and convenient mathematical calculations, as well as accessible experimental demonstrations, which are often deterministic in nature. Furthermore, one can also use CVs as a key element in another promising architecture, known as hybrid quantum information~\cite{Furusawa2011}.

The field of quantum computing~\cite{Ladd2010} using CVs~\cite{Lloyd1999,Weedbrook2011} has also progressed significantly in the last few years. From its original conception in 1999~\cite{Lloyd1999}, progress began to accelerate after a cluster state~\cite{Raussendorf2001} version was established in 2006~\cite{Zhang2006,Menicucci2006}, leading to something significantly more tangible for experimentalists. This resulted in numerous proof-of-principle demonstrations~\cite{Yokoyama2014,Miyata2014,Pysher2011,Takeda2013}, currently culminating in a 10,000 node cluster~\cite{Yokoyama2013} created `on-the-go' along with a 60 node cluster created simultaneously~\cite{Chen2014}. From a theoretical perspective, much progress has been made \cite{Marshall2014,Lau2013,Loock2007,Gu2009,Alexander2014,Demarie2014,Menicucci2015,Wang2014,Menicucci2011}, including recently, an important fault tolerant architecture~\cite{Menicucci2014}, achieved by leveraging the Gottesman-Kitaev-Preskill (GKP) encoding~\cite{Gottesman2001}. However, one area that is significantly underdeveloped is that of algorithms for a CV quantum computer. Thus far there only exists CV versions of quantum searching~\cite{Pati2000} and the Deutsch-Jozsa algorithm~\cite{Pati2003,Adcock2009,Zwierz2010,Adcock2013}.

In this paper, we present an algorithm that simulates~\cite{Georgescu2014} the scattering amplitudes in scalar bosonic quantum field theory (QFT) using a continuous-variable quantum computer. In fact, we show one can obtain an exponential speedup over the best known classical algorithms. A discrete version of the algorithm was originally shown in Refs.~\cite{Jordan2012,Jordan2014a} for a quantum computer based on qubits. Further work extended this result to fermionic QFTs~\cite{Jordan2014}, as well as using wavelets for multi-scale simulations~\cite{Brennen2014}. 

Typically, $q$ and $p$ are the CVs spreading across all real numbers. To encode them in qubits, one needs a whole register of qubits at each point in space. However, with CVs, there is a 1-to-1 mapping to qumodes (the CV equivalent of a qubit).  In fact it is arguable that a CV quantum computer is the natural choice for such a QFT problem given that​ the fields are continuous variables. Thus, the value of the field at a given point in space can be mapped onto a qumode naturally. If qubits are used, instead, the qumode needs to be replaced by a register of $M$ qubits which only allows the field to take on $2^M$ discrete values. Brennen \textit{et al.} describe both possibilities in Ref.~\cite{Brennen2014}, although they do not explain how to implement the quartic phase gate with CVs, which we do here. Furthermore, the quartic vertex in wavelets becomes very complicated. Implementing it would require gates acting on more than two modes (resulting in logarithmic overhead in complexity).  Another benefit to our approach is in the development of the initial cluster state. Here we show how to create the initial CV cluster state as well as suggesting an experimental implementation based on standard linear optics. Furthermore, we also note that in the preparation of the initial state we see a slight improvement over the original qubit approach of Ref.~\cite{Jordan2012}. There they require $O(N^{2.376})$ gates to engineer the ground cluster state; whereas in our scheme, we require slightly less than that, specifically, $O(N^2)$ gates. 

Our paper is structured in the following way. In Sec.~\ref{sec2}, we discretize space for a one-dimensional scalar bosonic QFT while leaving the field and time as continuous parameters.  Next, we show how to generate the initial cluster state using only Gaussian operations in Sec.~\ref{prep}.  In Sec.~\ref{compute} we outline the steps necessary to compute a scattering amplitude including the required measurement.  We provide an explicit experimental implementation in Sec.~\ref{experiment}.  Finally, the benefits of our approach over classical methods are discussed in Sec.~\ref{conclusion}.
\section{Discretization in one-dimension}
\label{sec2}

We consider a relativistic scalar field $\phi$ in one spatial dimension including a quartic self-interaction. We shall outline the discretization specifically in the one-dimensional case so as not to clutter the notation unnecessarily, but generalization to higher dimensions is straightforward and is discussed in the supplementary material. We note that the field $\phi$ is a function of $x$ and $t$ (time), $\phi(x,t)$. All three parameters are continuous. In our approach, we discretize $x$, but not $\phi$ or $t$. In the case of qubits, one would discretize $x$ and $\phi$, but not $t$. In classical lattice calculations, one discretizes all three $\phi$, $x$, and $t$.  

In the continuum, the one-dimensional free scalar QFT is given by the Hamiltonian
\be H_0 = \frac{1}{2} \int_0^L dx \left[ \pi^2 + \left( \frac{\partial\phi}{\partial x} \right)^2 + m^2 \phi^2 \right] \ee
where $\phi$ is the scalar field and $\pi$ the conjugate momentum field. They obey commutation relations $[ \phi (x),\pi (x') ] = i\delta (x-x')$
where we choose units in which $\hbar =1$.

We discretize space by letting $x = na$, $n=0,1,\dots, N-1$, where $a$ is the lattice spacing and $L=Na$ is the finite length of the spatial dimension ($L\gg a$). We choose units in which $a=1$, for simplicity, and denote $Q_n = \phi (x)$, $P_n = \pi(x)$. The discretized variables obey standard commutator relations, $[ Q_n,P_m ] = i \delta_{nm}$.
The Hamiltonian becomes
\be\label{eq18} H_0 = \sum_{n=0}^{N-1} \frac{ P_n^2 + m^2 Q_n^2}{2} +  \frac{1}{2} \sum_{n=0}^{N-1} (Q_n- Q_{n+1})^2 \ee 
where we employed periodic boundary conditions and defined $Q_{N} \equiv Q_0$.

It is useful to define  creation and annihilation operators, $A_n^\dagger$ and $A_n$, respectively, by ${A}_n = (Q_n + iP_n)/\sqrt{2}$.  They obey the commutation relations $[ A_n,A_m^\dagger ] = \delta_{nm}$ and the Hamiltonian can then be written as
\be \label{eq8} H_0 = \frac{1}{2} \mathbf{P}^T \mathbf{P} + \frac{1}{2} \mathbf{Q}^T \mathbf{V} \mathbf{Q} \ee
where $\mathbf{P} \equiv [ P_0,P_1, \dots, P_{N-1}]^T$ and $\mathbf{Q} \equiv [ Q_0, Q_1, \dots, Q_{N-1}]^T$ . The eigenvalues of the matrix $\mathbf{V}$ and the components of the corresponding normalized eigenvectors $\mathbf{e}^n$ are, respectively, $\omega_n^2 = m^2 + 4\sin^2 \frac{n\pi}{N}$, and $\mathbf{e}_{k}^{n} =  \frac{1}{\sqrt{N}} e^{2\pi i kn/N},\  k=0,\dots, N-1$.
Notice that the massless case is special because it contains a zero mode (for $m=0$, $\omega_0 =0$), so the matrix $\mathbf{V}$ is not invertible. To avoid the problems that arise, we can shift the mass by a small amount $\sim 1/N$, which vanishes in the continuum limit ($N\to\infty$).

We also wish to add a quartic interaction,
$ H_{int} = \frac{\lambda}{4!} \int_0^L dx \phi^4 \to  \frac{\lambda}{4!} \sum_n Q_n^4 $
which necessitates the addition of a mass counter term
$ H_{c.t.} = \frac{\delta_m}{2} \int_0^L dx \phi^2 \to \frac{\delta_m}{2} \sum_n Q_n^2 $
due to renormalization, as explained in the supplementary material.  We find that for weak coupling, the physically interesting case is stable for $\lambda>0$.

To diagonalize the Hamiltonian, we introduce new creation and annihilation operators, $a_k^\dagger$ and $a_k$, respectively, defined by ${a}_k = \sqrt{\frac{\omega_k}{2}} (\mathbf{e}^\dagger  \mathbf{Q})_k + \frac{i}{\sqrt{2\omega_k}} (\mathbf{e}^\dagger \mathbf{P})_k$
where $\mathbf{e}$ is the matrix of the eigenvectors. Notice that $ \mathbf{e}$ is unitary, $\mathbf{e}^\dagger \mathbf{e} = \mathbf{I}$.  These operators obey standard commutation relations, $[ a_k,a_l^\dagger ] = \delta_{kl}$ and the free Hamiltonian reads
\be H_0 = \sum_{k=0}^{N-1} \omega_k \left( a_k^\dagger a_k + \frac{1}{2} \right) .\ee
In this form, it is straightforward to construct the states in the Hilbert space.

\section{Initial cluster state preparation}\label{prep}

For the initial cluster state, in Refs.~\cite{Jordan2012,Brennen2014} the excited state was created \textit{after} creating the ground state. This is difficult because it involves manipulating a large number of qubits. In our approach, we create a single photon state in a single mode \textit{before} creating the cluster state. This is more accessible, as it involves creating the state $\ket{1}$ for a single mode. It can be done in a variety of ways, via a heralded single photon source, for instance. At the end of the computation, the field modes are all measured and the distribution of single photons across them determines the result.

To begin with, we build the system with $N$ oscillators representing the variables $(Q_n, P_n)$, $n=0,1,2,\dots$. The $n$th oscillator has a Hilbert space constructed by successive application of the creation operator $A_n^\dagger$ on the vacuum $|0\rangle_n$, which is annihilated by $A_n$. Here $|0\rangle_n$ is shorthand for a product state of vacuum fields
\be\label{eq16} |0\rangle = |0\rangle_0 \otimes |0\rangle_1 \otimes \cdots \otimes |0\rangle_{N-1} \ , \ee
with $ A_n |0\rangle = 0$.
For a scattering process, we are given an initial state typically consisting of a fixed number of particles, usually two, which undergoes evolution and then a measurement is performed (detection of particles) on the final state. Both initial and final states asymptote to eigenstates of the free Hamiltonian $H_0$. Thus quantum computation starts with preparation of an eigenstate of $H_0$.

\label{sec:groundstate}
First, we consider the ground state of $H_0$. It is the cluster state $|\Omega\rangle$ annihilated by all $a_k$, i.e., $ a_k |\Omega\rangle =0$ for $k=0,1,\dots, N-1 $.  It can be constructed from the vacuum state \eqref{eq16} by acting with the Gaussian unitary $U^\dagger$, where $a_n = U^\dagger A_n U $.  Noticing the relationship between the operators $a_k$ and $A_k$ we can use the Bloch-Messiah reduction \cite{Braunstein2005decomp} to determine $U=VSW^\dagger$ as a decomposition involving a multiport interferometer ($V$) followed by single mode squeezing ($S$) followed by a final multiport interferometer ($W$).    These unitary operators can be realized with $O(N^2)$ gates \cite{Reck1994}.  This is in contrast to the qubit version \cite{Jordan2012} where they require $O(N^{2.376})$ gates.

To implement $U$ we first perform the rotation 
\bes\label{eq23} A_0 &\to & A_0' = \sum_{k=0}^{N-1} A_k \nonumber\\
A_n &\to&  A_n' = \sum_{k=0}^{N-1} \cos \frac{2\pi nk}{N} A_k \nonumber\\
A_{N-n} &\to&  A_{N-n}' = \sum_{k=0}^{N-1} \sin \frac{2\pi nk}{N} A_k
\ees
where $1\le n\le N/2$,
which can be expressed in terms of rotations each involving only a couple of oscillators.
Notice that if $N$ is even, $A_{N/2}$ does not have a partner; we obtain $A_{N/2} \to \sum_k (-)^k A_k$.
Next, we squeeze each mode as $A_n' \to A_n'' = \cosh r_n A_n' + \sinh r_n {A_n'}^\dagger$
where $e^{2r_n} = \omega_n$ for $n\le N/2$, and $e^{-2r_n} = \omega_n$, for $n>N/2$.
Finally, we untangle the pairs by rotating them, $A''_k\to a_k$ where $a_0=A_0''$, $a_n =( A_n'' +  iA_{N-n}'' )/\sqrt 2$, and $a_{N-n} = ( iA_n'' +  A_{N-n}'' )/\sqrt 2$.
Excited states can be constructed with the same number of gates, e.g., the single-particle state $ |k\rangle \equiv a_k^\dagger |\Omega\rangle$ can be constructed by acting upon the vacuum with $A_k^\dagger$. This turns the initial state of the $k$th mode into a one-photon state, $A_k^\dagger |0\rangle_k$, which can be accomplished in a variety of ways; see supplementary material. Having engineered $ A_k^\dagger |0\rangle_k$, we then apply the Gaussian unitary $U^\dagger$, to obtain the one-particle state
\be {a_k}^\dagger |\Omega\rangle = U^\dagger A_k^\dagger |0\rangle \ee
Extending the above to the engineering of multi-particle states, $|k_1,k_2,\dots \rangle \propto {a_{k_1}}^\dagger {a_{k_2}}^\dagger \cdots |0\rangle$, is straightforward.

\section{Quantum Computation}\label{compute}

We wish to calculate a general scattering amplitude, which can be written as
\be \mathcal{A} = \langle out | T \exp\left\{ i \int_{-T}^T dt (H_{int} (t) + H_{c.t.} (t) ) \right\} | in \rangle \ee
in the limit $T\to\infty$, where time evolution is defined with respect to the non-interacting Hamiltonian.

We start by preparing the initial state $|in\rangle$ as in the previous section and define initial time as $t=-T$. Then we act successively with evolution operators of the form
\be\label{eq33}  U(t) = \exp\left\{ i \delta t (H_{int} (t) + H_{c.t.} (t) ) \right\}  \ee
Time dependence is obtained via the free Hamiltonian,
\be Q_i (t) = e^{it H_0} Q_i (0) e^{-it H_0} \ee
Therefore, the evolution \eqref{eq33} can be implemented as
\be\label{eq33a} U(t)  = e^{it H_0}e^{ i \delta t (H_{int} + H_{c.t.} ) } e^{-it H_0} \ee
We deduce
\be \mathcal{A} = \langle out | \left[ e^{i \delta t H_0} e^{ i \delta t (H_{int} + H_{c.t.} ) }  \right]^N |in\rangle \ee
where we divided the time interval into $N = \frac{2T}{\delta t}$ segments.

The coupling constants in \eqref{eq33} are turned on and off adiabatically. This is achieved by splitting the time interval $[-T,T]$ into three segments, $[-T, -T_1]$, $[-T_1, T_1]$, and $[T_1,T]$. For $t\in [-T,-T_1]$, we turn the coupling constants on by replacing $\lambda \to \lambda(t)$, $\delta m \to \delta m (t)$, so that $\lambda (-T) = \delta m (-T) =0$, and $\lambda (-T_1) = \lambda$, $\delta m (-T_1) = \delta m$. Then for $t\in [-T_1,T_1]$ the coupling constants are held fixed. Finally, for $t\in [T_1,T]$, they are turned off adiabatically by reversing the process in the first time interval. In the case of small $\lambda$, the time dependence of the coupling constants can be chosen efficiently by making use of perturbative renormalization. Eqs.\ \eqref{eqA13} and \eqref{eqA14} inform the choice $\lambda(t) = \frac{T+t}{T-T_1} \lambda$, $\delta m (t) = \frac{\lambda(t)}{8\pi} \log \frac{64}{m^2}$, for $-T \le t \le -T_1$.

The unitary operators $e^{i\delta t H_0}$ and $e^{i\delta t H_{c.t.}}$ are Gaussian and can be implemented with second order nonlinear optical interactions and linear optics beam splitter networks. The interaction is implemented through a \emph{quartic} phase gate for each mode,
\be\label{eq41} e^{i\delta t H_{int} } = \prod_n e^{i\gamma Q_n^4 } \ \ , \ \ \ \ \gamma = \delta t \frac{\lambda}{4!}  \ee
The quartic phase gate may be implemented in a similar manner to the cubic phase gate previously proposed~\cite{Marshall2014}.

\begin{figure*}[t]
\includegraphics[width=6.5in]{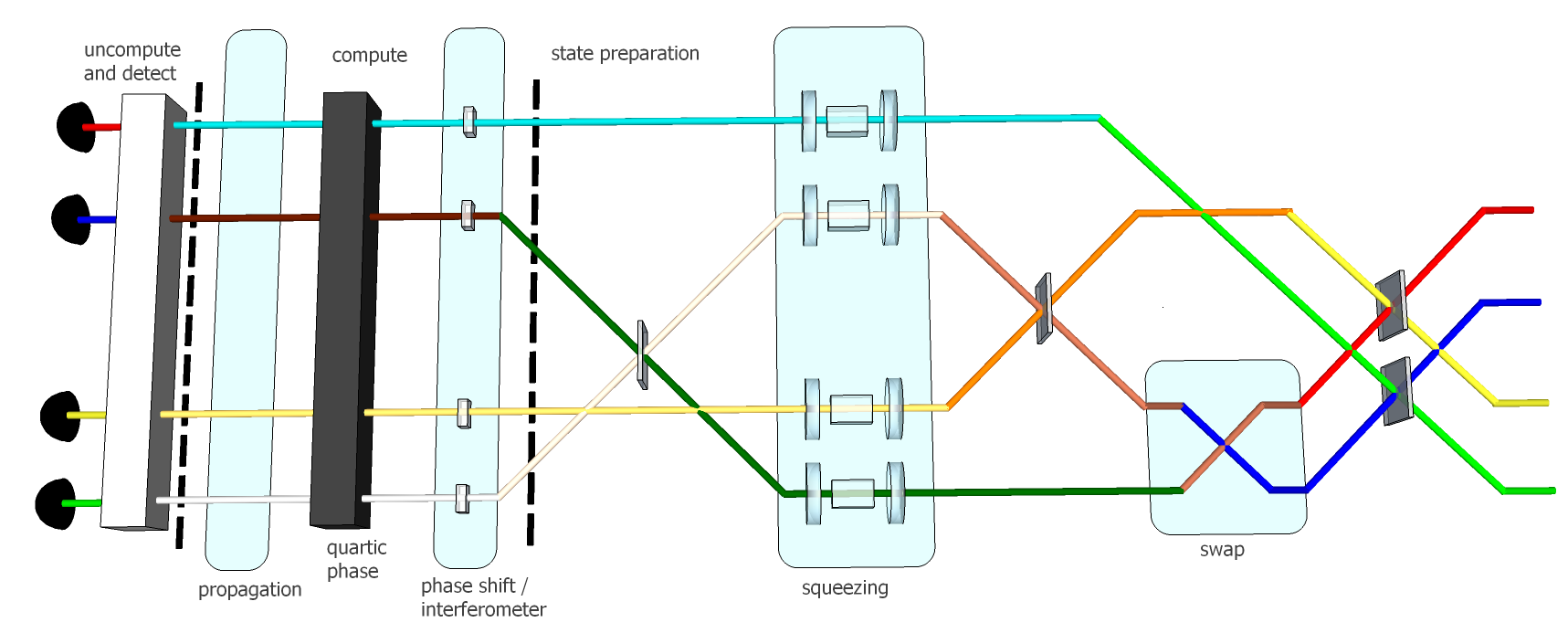}
\captionsetup{justification=raggedright,
singlelinecheck=false
}
\caption{(Color Online) Sketch of an experimental setup for electromagnetic field modes used as qudits in a QFT calculation involving four field modes. The modes are encoded into electric field modes (colored red, blue, yellow, green), which are then prepared via beam splitters, swap gates, and squeezers for the compute stage. The compute stage consists of an interferometer, a quartic phase gate (black box, see Ref.~\cite{Marshall2014}), and free propagation. An uncompute stage, which is the inverse of the preparation stage, and a detection stage in the Fock basis, yield the scattering amplitudes into the four QFT field modes.}\label{setup}
\end{figure*}

After evolution, we must project onto the state $|out\rangle$. This is similar to the state $|in\rangle$, and its construction depends on the number of desired particles. The latter are excitations created with $a_n^\dagger$, so  in general,
\be\label{eq95} |out\rangle = a_{n_1}^\dagger a_{n_2}^\dagger \cdots |\Omega\rangle = U^\dagger A_{n_1}^\dagger A_{n_2}^\dagger \cdots |0\rangle\ee
It follows that the next step is to \emph{uncompute} by applying the Gaussian unitary $U$ (which is the inverse operation to the preparation of the initial state), and then measure the number of photons in each mode. The final uncompute step projects the set of output modes onto the Fock basis. Thus, the scattering amplitude calculation is a mapping from one set of field modes on the input to a separate set of field modes on the output, as expected.
That is, for each click on the photodetector for mode $n$, there is an operator $a_n^\dagger$ present in the final state \eqref{eq95}. If the QFT calculation involved an initial input state with two excitations spread across 100 field modes, say, then the entire calculation would involve two photons, for instance. We note that the calculation has made use of a quartic phase gate up to this point, and thus technically speaking a non-Gaussian operation would not be necessary during this measurement step in order to achieve an exponential speedup over the classical QFT algorithm. However, in order to achieve high accuracy in the final result, photon number resolving detectors with high efficiency~\cite{Nam2008} would be desirable for the measurment phase. 

 
\section{Experimental implementation}\label{experiment}

An example of the experimental implementation is given in Fig.~\ref{setup}. For brevity the  setup for calculating four space time points is given. For the electromagnetic field, the initial unitary rotation involves weighted beam splitters with the appropriate splitting to achieve the desired sums over the field operators (see appendix A.2). A swap gate is involved in the input state preparation stage. We note that a swap gate contains essentially the CV version of the CNOT operator along with parity operators~\cite{Wang2001}, but in some cases the gate can be simplified to a beam splitter interaction~\cite{Braunstein2005} such as for the electromagnetic field. Here we use a mode label swap operator, which is possible in systems with movable qubits, such as CV optical fields.
Next, $H_{c.t.}$ is quadratic in position quadrature operators, which can be implemented with a series of phase shifts~\cite{Braunstein2005}. The non-Gaussian piece of the computation is then the quartic phase gate contained in $H_{int}$, which can be implemented via repeated application of the photon number-dependent phase gate~\cite{Marshall2014}. Lastly, the free propagation $H_0$ can be implemented by a calibrated free propagation before the uncompute stage.  We note that the QFT field modes are encoded into the qudits which are themselves electromagnetic field modes, meaning that the free propagation contained in $H_0$ is not arbitrary. It must conform to the calculated QFT free propagation distance, and phase stability must be maintained throughout.

\section{Conclusion}\label{conclusion}

In conclusion, we developed a new algorithm for a continuous-variable quantum computer which gave an exponential speedup over the best known classical algorithms. This algorithm was the calculation of the scattering amplitudes in scalar bosonic quantum field theory, and as previously mentioned, arguably a natural choice for a continuous variable quantum computer to solve.  At weak coupling, analytic calculations are possible, however, at strong coupling no such calculations are generally available, and one has to rely on numerical techniques. A widely used framework is lattice field theory which is based on the discretization of space into a finite set of points. The complexity of classical computations on a lattice increases exponentially with the number of lattice sites \cite{Creutz1985}.  

Quantum computations offer a distinct advantage (first shown in Ref.~\cite{Jordan2014a} for qubits, and here for qumodes), since complexity only grows polynomially.  Using continuous variables we also see an advantage over the original qubit proposal; specifically, in the preparation of the initial cluster state. There they required $O(N^{2.376})$ gates to synthesize the ground state. However, in our scheme, we required slightly less, $O(N^2)$ gates. Finally, we also gave an example of an experimental implementation on a continuous-variable cluster state quantum computer that calculated four space time points. We noted that such a scheme is feasible with current linear optical technology and consisted of a set of Gaussian operations along with the non-Gaussian quartic phase gate.

\acknowledgments We are grateful to Peter Rohde for valuable feedback.  R. C. P. performed portions of this work at Oak Ridge National Laboratory, operated by UT-Battelle for the US Department of Energy under Contract No. DE-AC05-00OR22725. Work performed by the US government is not subject to copyright restrictions.

\appendix
\section{Renormalization}

Define the Green function $\mathbf{G} (t_1,t_2)$ as 
\be G_{ij}(t_1,t_2) = \langle 0 | \mathcal T(Q_i (t_1) Q_j(t_2)) |0\rangle, \ee where $\mathcal T$ denotes the time-ordering operator.  It obeys \be\left[  \partial_{t_1}^2 + \mathbf{V} \right] \mathbf{G} (t_1,t_2) = -i \mathbf{I} \delta (t_1-t_2).\ee  Using the Fourier transform, \be \mathbf{G} (t_1,t_2) = \int \frac{d\omega}{2\pi} e^{i\omega (t_1-t_2)} \tilde{\mathbf{G}} (\omega)\ee
we obtain
\be \tilde{\mathbf{G}} (\omega) = i\left[ -\omega^2 \mathbf{I} + \mathbf{V} \right]^{-1} 
= \sum_n \frac{-i}{\omega^2 - \omega_n^2} \mathbf{e}_n \mathbf{e}_n^\dagger, \ee
exhibiting poles at $\omega^2 = \omega_{n}^2$. 

When we switch on the interaction term, \be H_{int} = \frac{\lambda}{4!} \int_0^L dx \phi^4 \to  \frac{\lambda}{4!} \sum_n Q_n^4,\ee we have that at $\mathcal{O} (\lambda)$ the Green function is corrected by
\be\delta G_{ij}(t_1,t_2) =  \langle 0 | \mathcal T\left[ Q_i (t_1) Q_j(t_2) \int dt  H_{int} (t) \right] |0\rangle.\ee
For the Fourier transform, we obtain \be\delta \tilde{\mathbf{G}} (\omega) = \lambda [\tilde{\mathbf{G}} (\omega)]^2 \int \frac{d\omega'}{2\pi} \mathrm{Tr}\, \tilde{\mathbf{G}} (\omega')\ee
which leads to a shift of the poles,
\be  \tilde{\mathbf{G}} (\omega) + \delta \tilde{\mathbf{G}} (\omega) = \sum_n \frac{-i}{\omega^2 - \omega_n^2 - \Sigma} \mathbf{e}_n \mathbf{e}_n^\dagger + \mathcal{O} (\lambda^2), \ee
where
\be \Sigma = \frac{\lambda}{2N} \int \frac{d\omega'}{2\pi} \sum_n \frac{-i}{{\omega'}^2 -\omega_i^2} = \frac{\lambda}{4N} \sum_n \frac{1}{\omega_n} \ee
The shift can be corrected by the addition of the counter term \be H_{c.t.} = \frac{\delta_m}{2} \int_0^L dx \phi^2 \to \frac{\delta_m}{2} \sum_n Q_n^2,\ee with $\delta_m = -\Sigma + \mathcal{O} (\lambda^2)$, i.e., the mass parameter in the Hamiltonian is not physical, but bare,
\be m_0^2 = m^2 + \delta_m = m^2 - \frac{\lambda}{4N} \sum_n \frac{1}{\omega_n} + \mathcal{O} (\lambda^2). \ee
For large $N$, the sum can be approximated by an integral,
\be \Sigma = \frac{\lambda}{4} \int_0^1 \frac{dk}{\sqrt{m^2 + 4\sin^2 k\pi}} \ee
which has a logarithmic divergence at small $m^2$ (i.e., length scale $1/m$ large in units of lattice spacing, which is the physically interesting limit).
We easily obtain
\be \label{eqA13}\Sigma = \frac{\lambda}{8\pi} \log \frac{64}{m^2} + \mathcal{O} (m^2) \ee
The bare mass is
\be \label{eqA14} m_0^2 = m^2 -\Sigma + \mathcal{O} (\lambda^2) = m^2 - \frac{\lambda}{8\pi} \log \frac{64}{m^2} + \mathcal{O} (\lambda^2, m^2) \ee
Notice that for weak coupling (small $\lambda$), the physically interesting case has $m_0^2 < 0$ (a stable system, as long as $\lambda >0$).

\subsection{Ground State Construction}
To find the required transformation $U$, we work as follows. Notice that for $n=0$,
\be a_0 = \frac{1}{2\sqrt{N}} \sum_{k=0}^{N-1}\left[ \left( \sqrt{m} + \frac{1}{\sqrt{m}} \right)  A_k +  \left( \sqrt{m} - \frac{1}{\sqrt{m}} \right) A_k^\dagger \right] \ee 
where we used $\omega_0 =m$. For $n\ne 0$, we consider pairs $(a_n , a_{N-n})$. We have
\bes a_n + a_{N-n} &=& \frac{1}{2\sqrt{N}} \sum_{k=0}^{N-1} \cos \frac{2\pi nk}{N}
\left[ \left( \sqrt{\omega_n} + \frac{1}{\sqrt{\omega_n}} \right)  A_k \right. \nonumber\\
&& \left. +  \left( \sqrt{\omega_n} - \frac{1}{\sqrt{\omega_n}} \right) A_k^\dagger \right] \nonumber\\
a_n - a_{N-n} &=& \frac{i}{2\sqrt{N}} \sum_{k=0}^{N-1} \sin \frac{2\pi nk}{N}
\left[ \left( \sqrt{\omega_n} + \frac{1}{\sqrt{\omega_n}} \right)  A_k \right. \nonumber\\
&& \left. -  \left( \sqrt{\omega_n} - \frac{1}{\sqrt{\omega_n}} \right) A_k^\dagger \right]
\ees
where we used $\omega_n = \omega_{N-n}$.

The above expressions suggest that we transform $A_n$ into $a_n$ in three steps as detailed in Sec. \ref{sec:groundstate}

\subsection{Example: $N=4$}

To illustrate the above algorithm,  we consider the case in which space has been discretized to four points. The rotation  ($\mathbf{A}' = \mathbf{O} \mathbf{A}$) is described by the orthogonal matrix
\be \mathbf{O} = \frac{1}{2} \left[ \begin{array}{cccc} 1 & 1 & 1 & 1 \\ \sqrt{2} & 0 & -\sqrt{2} & 0 \\ 1 & -1 & 1 & -1 \\ 0 & \sqrt{2} & 0 & -\sqrt{2} \end{array} \right] \ee
We have
\be \mathbf{O} = R_{02} \left( \frac{\pi}{4} \right) S_{01} R_{13} \left( \frac{\pi}{4} \right)  R_{02} \left( \frac{\pi}{4} \right) \ee
where $R_{ij} (\theta)$ is a rotation in the $ij$-plane of angle $\theta$ and $S_{ij}$ is the swap $i\leftrightarrow j$. Therefore the rotation $\mathbf{O}$ can be implemented with four two-mode unitaries.

Next, we squeeze each mode as
$ A_n' \to A_n'' = \cosh r_n A_n' + \sinh r_n {A_n'}^\dagger$, where $e^{2r_0} = \omega_0$, $e^{2r_1} = \omega_1$, $e^{2r_2} = \omega_2$, and $e^{-2r_3} = \omega_3$. Notice that $r_3 = - r_1$, because $\omega_3 = \omega_1$.

Finally, we perform the rotation, $A_1'' \to  \frac{1}{\sqrt{2}} ( A_1'' +iA_3'')$, $A_3'' \to \frac{1}{\sqrt{2}} (iA_1'' + A_3'')$, to arrive at the desired modes,
\bes a_0 &=& \frac{1}{2} \sum_n \left[ \cosh r_0 A_n  + \sinh r_0 \sum_n A_n^\dagger \right] \nonumber\\
 a_1 &=& \frac{1}{2} \sum_n i^n \left[ \cosh r_1 A_n  + \sinh r_1 \sum_n A_n^\dagger \right] \nonumber\\
 a_2 &=& \frac{1}{2} \sum_n (-1)^n \left[ \cosh r_2 A_n  + \sinh r_2 \sum_n A_n^\dagger \right] \nonumber\\
 a_3 &=& \frac{1}{2} \sum_n (-i)^n \left[ \cosh r_3 A_n  + \sinh r_3 \sum_n A_n^\dagger \right]
\ees

Each of the above steps is implemented with a Gaussian unitary involving at most two modes.

\section{Excited States}
To generate the required one-photon state, two methods can be used. One can first squeeze the vacuum of the $k$th mode with an optical parametric amplifier to
\be S_k(s) |0\rangle_k \ \ , \ \ \ \ S_k(s) = e^{\frac{s}{2} ({A_k^\dagger}^2 - A_k^2 )} \ee
Then pass the squeezed state through a (highly transmitting) beam splitter of transmittance $T$, and place a photodetector on the auxiliary output port. A click of the detector heralds a successful photon subtraction, which is described by the non-unitary operator
\be \sqrt{1-T}\, T^{A_k^\dagger A_k/2}  A_k \ee
The transmittance has to be high so that the probability of detecting two or more photons is negligible. If no photon is detected, the process is repeated until a photon is detected.
Finally, apply anti-squeezing $S_k^\dagger (s')$.

We obtain the state (unnormalized)
\be\label{eq72} S_k^\dagger (s') T^{A_k^\dagger A_k/2}  A_k S_k(s) |0\rangle_k \ee
If the squeezing parameters are chosen so that
\be T = \frac{\tanh s'}{\tanh s} \ee
then it is straightforward to show that \eqref{eq72} is the desired state,
\be S_k^\dagger (s') T^{A_k^\dagger A_k/2}  A_k S_k(s) |0\rangle_k \propto A_k^\dagger |0\rangle_k. \ee

Optionally, one may also use a heralded single photon source. Such a source would consist of a parametric downconverter with a high efficiency heralding detector. To obtain exactly one photon when operating the source with high brightness (but on average less than one pair per pulse), the heralding detector would consist of a high efficiency photon number resolving detector, such as a transition edge sensor.

\section{Generalization to Arbitrary Dimensions}

Generalization to arbitrary spatial dimension $d$ is straightforward. The free-scalar Hamiltonian in the continuum reads
\be\label{eq61} H_0 = \frac{1}{2} \int d^dx \left[ \pi^2 + (\mathbf{\nabla} \phi )^2 + m^2 \phi^2 \right] \ee
where $\mathbf{x} \in [0,L]^d$, with the fields obeying standard commutation relations,
\be [ \phi ( \mathbf{x} ) \ , \ \pi(\mathbf{x}') ] = i \delta^d (\mathbf{x} - \mathbf{x}' ) \ee
Each coordinate $x_i$ ($i=1,\dots, d$) is discretized as before, $x_i = n_i a$, $n_i = 0,1,\dots, N-1$, and we define $Q_{\mathbf{n}} \equiv \phi (\mathbf{x})$, $P_{\mathbf{n}} \equiv \pi (\mathbf{x})$, $A_{\mathbf{n}} = \frac{1}{\sqrt{2}} (Q_{\mathbf{n}} + iP_{\mathbf{n}})$, where $\mathbf{n} \in \mathbb{Z}_N^d$.

The Hamiltonian \eqref{eq61} can then be rendered in the form \eqref{eq8}, where $\mathbf{V}$ has eigenvalues and corresponding normalized eigenvectors,
\bes \omega_{\mathbf{k}}^2 &=& m^2 + 4 \sum_{i=1}^d \sin^2 \frac{k_i}{2} \ , \nonumber\\
\mathbf{e}_{\mathbf{k}}^{\mathbf{n}} &=& \frac{1}{N^{d/2}} e^{i\mathbf{k} \cdot \mathbf{n}} \ees
where $\mathbf{k} \in \frac{2\pi}{N} \mathbb{Z}_N^d$ (the dual lattice). The eigenvectors form a unitary matrix.

The discretized Hamiltonian is diagonalized as
\be H_0 = \sum_{\mathbf{k} \in \Gamma} \omega_{\mathbf{k}} \left( a_{\mathbf{k}}^\dagger a_{\mathbf{k}} + \frac{1}{2} \right) \ee
where $a_{\mathbf{k}}$ is the annihilation operator defined in Sec.~\ref{sec2} (extended to $d$ dimensions in an obvious way).

Introducing an interaction term, $H_{int} = \frac{\lambda}{4!} \sum_{\mathbf{n}} Q_{\mathbf{n}}^4$, and the attendant counter term, $H_{c.t.} = \frac{\delta_m}{2} \sum_{\mathbf{n}} Q_{\mathbf{n}}^2$, and working as in the one-dimensional case, we obtain a shift in the poles of the Green function,
\be \Sigma = \frac{\lambda}{4} \sum_{\mathbf{k} \in \Gamma} \frac{1}{\omega_{\mathbf{k}}} + \mathcal{O} (\lambda^2) \ee
which is related to the counter-term parameter $\delta_m$ via $\delta_m = -\Sigma + \mathcal{O} (\lambda^2)$.  For large $N$, the sum is approximated by an integral over the hypercube $[0,2\pi]^d$.  For $d=1$, it reduces to the previous result, whereas for $d>1$, we obtain at lowest order in $m$ and $\lambda$,
\be \Sigma = C_d \lambda + \dots \ee
Numerically, $C_2 \approx 0.16$, and $C_3 \approx 0.11$.


\end{document}